\documentstyle[12pt]{article}

\begin{document}
\noindent
\begin{center}
\begin{large}
{\bf Conditional calculus on fractal structures and\\
 its application to galaxy distribution}\\
 \end{large}
\end{center}
\vspace{1cm}
\vspace{1cm}
\begin{center}

Yu Shi\\
Department of Physics, Bar-Ilan University, Ramat-Gan 52900, Israel
\\ 
\end{center}
\vspace{1.5cm}


\newpage
{\bf Abstract}
It is shown that calculus can apply on a fractal structure
with the condition that the infinitesimal  limit of 
change of the variable is larger than the lower cut-off of the 
fractal structure,
and an assumption called local decomposability.
As an application, it is shown  that the angular
 projection of a fractal distribution in 3-dimensional space
 is not homogeneous at sufficiently large angles. Therefore
 the angular projection of galaxy distribution for sufficiently
 large angles can discriminate the fractal and the homogeneity
 pictures.

\newpage

It seems that calculus cannot apply in a fractal structure, which
possesses infinite places of discontinuity. However, it is shown in this letter
that it can be done in a conditional way. Just as that only with
 calculus can we deal with manifold and
function with an arbitrary shape, the conditional calculus on fractals will enable
us deepen the studies  of fractal. 

This work is partly motivated by
the investigations on fractal distribution of  galaxies.
 While it is an agreement that the galaxy distribution approximates
  a fractal  over a considerable range of scales,
   it is  under  a 
   debate whether it becomes  homogeneous on a scale about, say, 
   $20h^{-1}Mpc$ \cite{Borgani}
  \cite{Davis}, or is fractal up to the present
  observational limit 
 \cite{Pietronero}\cite{CP}. A 
  crucial problem, which is also  controversial,
  related to
   the above question is whether the angular
   projection of a fractal embedded in a 3-dimensional (3d) 
    space  is homogeneous. Since before the extensive redshift survey the
    galaxy catalogs were for 
    angular coordinates and now there 
    is much angular information  \cite{Davis},
    a difference in angular projection will help us choose between the
    two alternative pictures. 
     Based on a numerical simulation on a fractal structure geneated by a
 Levy flight in 3d, 
 it was stated that the angular projection shows 
 power-law correlation at small 
 angles  but becomes homogeneous at large angles
 \cite{CP}.  However, as we pointed out in a previous work \cite{Shi}, 
 their explaination 
  is inconsitent, taking 
 the implicit assumption of angular 
 homogeneity in distribution. We distinguished the concepts of  
angular distribution and angular projection, the former refers to the number
of points as a function of polar 
angle within a given radial depth while the latter
is defined through the solid-angular density represented as a function of
polar angle. It was shown that the power-laws for angular projection at
small angles are manifestinations of the
 so-called local angular fractal, which refers to that for a small
 conic part of a sphere defined by the angle $2\theta$ and the depth $L$,
  when $\theta$ is small enough, the points on each spherical shell within 
  the conic part have fractal distribution. The local angular fractal
  may reconcil the evidence of fractal and that claimed to be 
  of homogeneity, and 
  realize Cosmological Principle in a fractal structure in a rather
  strict way.
  The behavior at large
  angles remained open there, and will be obtained here through the 
  application of conditional calculus. Methods of calculus is also 
  expected when one deals with the fractal structure on a curved space, 
  which should 
  be taken into account for large-scale structure of the universe.
  
To define a function on a fractal, there are two alternative sets 
of independent variables, one is the continuous variables, 
or in geometric interpretation, coordinates of the continuous space
on which the fractal is embedded,
 another is the variables or coordinates restricted on the
fractal,  the inner coordinates,
effectively they form a continuous subspace. A fractal is
only a fractal looked in the embedding space, so we do not know whether
the latter viewpoint is useful, practically one takes the 
former.

 Let us start with the derivative of $f(x)$, $f'(x)\,=\,
\lim_{\Delta x\rightarrow 0} [f(x+\Delta x)-f(x)]/\Delta x$. For it to be
meaningful, $f(x+\Delta x)$ should be within the fractal set, therefore 
$\Delta x$ should conditionally approach $0$ so that it is larger than the
lower cut-off size of the fractal structure.  Therefore we have
the following condition of infinitesimal.

{\em Condition. The infinitesimal limit of change of variable
 should be taken in such a 
 way that it is larger than the lower cut-off size of the fractal 
structure.}

A mathematically-ideal fractal is infinitely iterated and 
there is no lower cut-off,  the above
 condition is automatically satisfied. Practically, this means 
 that the fractal under study 
should be large enough.
 The derivative on a fractal has been used, e.g.,
 in defining the conditional density \cite{CP}, where the above condition
  is actually an implicit assumption. So there is almost
  no difficulty in using
  derivative.
            
The difficulty lies in integral. How can we integrate a function $f(x)$
only
over the points belonging to the fractal? In the case of 1-dimensional
 embedding
space, it can be done through change of variable. Suppose
$x_{f}\,=\,
c_{x} x^{D_{x}}$, where the subscript ``f'' denotes the variable
on the fractal, $c_{x}$ is a coefficient and $D_{x}$ is 
the fractal dimension, then  $dx_{f}\,=\,c_{x}D_{x}x^{D_{x}-1}dx$,
 the integral is thus 
\begin{equation}
I\,=\,\int_{x_{f}(a)}^{x_{f}(b)}f(x)dx_{f}\,=\,c_{x}D_{x}\int_{a}^{b}
f(x)x^{D_{x}-1}dx,
\end{equation}
where $a$ and $b$ are the limits of the integration.

Now we entend it to multiple integral, which can only be done by
 transformed to an iterated one. In the ordinary case of
 continuous space
 this is  based on expressing the region of integration in terms of
  the infinitesimal
 changes of the  independent variables, in geometric interpretation,
 the element of volume is expressed in the infinitesimal changes of
 curvilinear coordinates. On a fractal, the element of volume can be
 expressed as the product of the variables on the fractal, then transformed
 to those of the embedding continuous space. Here the
 following assumption is needed. 

 {\em Assumption. The infinitesimal element of volume on a fractal is
 a product of infinitesimal changes of lengths in orthogonal directions, 
 these infinitesimal changes are 
 fractal subsets of the infinitesimal element of volume.}

This can be called the assumption of local decomposability.
 It can be seen that this assumption is based on the condition of 
 infinitesimal. It is most likely that under this condition
 the above assumption can be valid, especially in many random fractals
 such as that generated by a Levy flight and the galaxy distribution
  on the scales exhibiting fractal structure.
    The sum of the dimensions of the fratal subsets
  is, of course, the total dimension of the fractal. But we have no reason
that the the decomposition is independent of the position, the dimensions of
the subfractal may depend on position. If they are independent of the
 position, the fractal can be referred to as being uniform.

 Thus the multiple integral on a fractal can be done as an iterated integral
 over the subfractals, with the variables then 
 changed to those of embedding continuous
 space. For example, the volume integral in terms of 
 the Cartesian coordinates is
 $\int\int\int_{V} f(x)dV_{f}\,=\,\int \int\int f(x)dx_{f}dy_{f}dz_{f}\,=\,
c_{x}c_{y}c_{z}\int \int\int f(x)x^{D_{x}-1}y^{D_{y}-1}z^{D_{z}-1}dxdydz$.
In this way all  calculus can apply to fractals.

Formally, calculus may apply with the variables of fractals by 
transforming derivative with the  continuous variable  to that 
with  the inner variable of the fractal, {\it i.e.}, 
$df(x_{f})/dx_{f}\,=\,c_{x}^{-1}x^{1-D_{x}}df[x_{f}(x)]/dx$. For expample,
this  derivative of number of points is a constant, showing the fractal
subset form an effective homogeneous space. Seen from the embedding space,
this derivative is fractional. 

Now we turn to the angular projection of a conic part defined by
$L$ and $2\theta$, of a 
fractal in 3-dimensional space.  In terms of spherical coordinates
$(r,\theta,\phi)$, the
number of points in this volume is 
\begin{eqnarray}
N(L,2\theta)&=&\int_{0}^{L}\int_{0}^{\theta}\int_{0}^{2\pi}
\partial_{r}(c_{r}r^{D_{r}})\partial_{\theta}
[c_{\theta}(r\theta)^{D_{\theta}}]
\partial_{\phi}[c_{\phi}(r\sin\theta \phi)^{D_{\phi}}]
\nonumber\\
 &=& AL^{D}\cdot
 \int_{0}^{\theta}\theta^{D_{\theta}-1}(\sin\theta)^{D_{\phi}}
 d\theta,
 \label{eq:rtp}
 \end{eqnarray}
 where   $A\,=\,c_{r}c_{\theta}c_{\phi}D_{r}D_{\theta}(2\pi)^{D_{\phi}}$,
$D_{r}$, $D_{\theta}$, $D_{\phi}$ and $D$ are fractal dimensions of the 
infinitesimal subfractals and the total fractal, respectively.
$\partial_{r}$,  $\partial_{\theta}$ and $\partial_{\phi}$ represent 
partial derivatives.  
The second equality  is valid when  $D_{r}$, $D_{\theta}$ and $D_{\phi}$
are constants.  When $\theta$ is small, we obtain $N(L,2\theta)\,=\,
AL^{D}\theta^{D_{\theta}+D_{\phi}}$, which is just the ``local
angular fractal'' discussed in \cite{Shi}. If an arbitrary azimuthal angle
$\phi$ is considered, the number of points are then proportional to
 $\phi^{D_{\phi}}$. We suggest this  be tested for galaxy distribution.  

For an isotropic fractal, all directions are equivalent each other, so
the dimension of the subfractal in each direction is $D/3$, i.e. 
$D_{r}\,=\,D_{\theta}\,$$=\,D_{\phi}\,$$=\,D_{x}\,=\,D_{y}$$
\,=\,D_{z}\,=\,D/3$. This is most likely satisfied by the galaxy distribution
where $D\,\approx\,2$ while $D_{\theta}\,\approx\,1.3$ \cite{Pietronero}
\cite{Shi}.

 The solid angle is $2\pi(1-\cos \theta)$, so 
 the conditional angular density defined from the origin
  is $\Gamma(\theta)\,=\,dN(\theta)/d\Omega(\theta)\,=\,$
 $(dN/d\theta)/(d\Omega/d\theta)\,=\,$
 $(A/2\pi)L^{D}F(\theta)$, where
 $F(\theta)\,=\,\theta^{D_{\theta}-1}(\sin\theta)^{D_{\phi}-1}$.
 If the sample size is characterized by  $2\theta_{M}$ and  $L$,
   the average angular density over the sample is 
 $<n>_{M}\,=\,N(\theta_{M})/\Omega(\theta_{M})$,
 which is surely dependent on $\theta_{M}$.
 The angular correlation function is 
 $\omega(\theta)\,=\,\Gamma(\theta)/<n>_{\theta_{M}}-1$, which
 is dependent on $\theta$ through $\Gamma(\theta)$, and also on $\theta_{M}$. But 
 since it breaks down spuriously at angles much smaller than 
 $\theta_{M}$ \cite{CP}\cite{Shi}, the dependence of angular projection
 on $\theta$ should be examined in $\Gamma(\theta)$, thus just $F(\theta)$.

 The  dependence of $F(\theta)$ on $\theta$ can be
  seen from Figure 1, there is a power-law region at small angles, then 
  there is a relatively flat region, which is not so short since the plot is 
  a  log-log one, but finally it increases with $\theta$. So the conclusion
is that the angular projection of fractal is not homogeneous. 

For the fractal generated by a Levy flight \cite{CP}
is isotropic implied by the generation
rule. For the galxy distribution, there is also much evidence of isotropy 
\cite{Davis}. So the dimension of infinitesimal 
subfractals  are likely to be indeed 
independent on the position (cf. discussions in \cite{Shi}).
 Even if they are dependent on position,
the qualitative nature of 
dependence of the conditional angular  density on the angle is impossible to 
change. Though we obtain the result disagreeing with the claim
in
 \cite{CP},
 it is not
contradictory with what was actually reported, 
 only a very small region was given there,
 and  after a fairly flat region of 
 $\sim 0.3^{\circ}$, there can be observed 
  an indication of increase. So the claimed
 homogeneity of angular projection
 at large angles is quite 
 unsure from there.

So the behavior of
conditional angular density at sufficiently large angles can discriminate
fractal model and homogeneous model for the galaxy distribution on the
corresponding scale. Unfortunately, to our knowledge, there is still no such 
evidence. It should be noted that one should use $\Gamma(\theta)$ instead of
$\omega(\theta)$ since $\omega(\theta)$ decreases rapidly at angles
much smaller than  $\theta_{M}$. 
Among the interesting problems also is that to 
investigate in various fractal structures and the galaxy distribution whether
the assumption of local decomposability is indeed valid, and the dependence
or independence of the dimensions of local subfractals on the position.

\newpage
Figure Caption:

{\bf Figure 1.}  Log-log plot of $F(\theta)\,=\,\theta^{D_{\theta}-1}$
$(\sin\theta)^{D_{\phi}-1}$ by setting $D_{\theta}\,=\,0.6$ 
and $D_{\phi}\,=\,0.7$ so
that $D_{\theta}+D_{\phi}$ equals what was estimated for galaxy 
distribution \cite{Shi}.

\newpage
\begin{figure}
% GNUPLOT: LaTeX picture
\setlength{\unitlength}{0.240900pt}
\ifx\plotpoint\undefined\newsavebox{\plotpoint}\fi
\begin{picture}(1500,900)(0,0)
\font\gnuplot=cmr10 at 10pt
\gnuplot
\sbox{\plotpoint}{\rule[-0.200pt]{0.400pt}{0.400pt}}%
\put(220.0,113.0){\rule[-0.200pt]{2.409pt}{0.400pt}}
\put(1426.0,113.0){\rule[-0.200pt]{2.409pt}{0.400pt}}
\put(220.0,130.0){\rule[-0.200pt]{2.409pt}{0.400pt}}
\put(1426.0,130.0){\rule[-0.200pt]{2.409pt}{0.400pt}}
\put(220.0,144.0){\rule[-0.200pt]{2.409pt}{0.400pt}}
\put(1426.0,144.0){\rule[-0.200pt]{2.409pt}{0.400pt}}
\put(220.0,156.0){\rule[-0.200pt]{2.409pt}{0.400pt}}
\put(1426.0,156.0){\rule[-0.200pt]{2.409pt}{0.400pt}}
\put(220.0,167.0){\rule[-0.200pt]{2.409pt}{0.400pt}}
\put(1426.0,167.0){\rule[-0.200pt]{2.409pt}{0.400pt}}
\put(220.0,177.0){\rule[-0.200pt]{4.818pt}{0.400pt}}
\put(198,177){\makebox(0,0)[r]{0.1}}
\put(1416.0,177.0){\rule[-0.200pt]{4.818pt}{0.400pt}}
\put(220.0,241.0){\rule[-0.200pt]{2.409pt}{0.400pt}}
\put(1426.0,241.0){\rule[-0.200pt]{2.409pt}{0.400pt}}
\put(220.0,278.0){\rule[-0.200pt]{2.409pt}{0.400pt}}
\put(1426.0,278.0){\rule[-0.200pt]{2.409pt}{0.400pt}}
\put(220.0,305.0){\rule[-0.200pt]{2.409pt}{0.400pt}}
\put(1426.0,305.0){\rule[-0.200pt]{2.409pt}{0.400pt}}
\put(220.0,325.0){\rule[-0.200pt]{2.409pt}{0.400pt}}
\put(1426.0,325.0){\rule[-0.200pt]{2.409pt}{0.400pt}}
\put(220.0,342.0){\rule[-0.200pt]{2.409pt}{0.400pt}}
\put(1426.0,342.0){\rule[-0.200pt]{2.409pt}{0.400pt}}
\put(220.0,356.0){\rule[-0.200pt]{2.409pt}{0.400pt}}
\put(1426.0,356.0){\rule[-0.200pt]{2.409pt}{0.400pt}}
\put(220.0,368.0){\rule[-0.200pt]{2.409pt}{0.400pt}}
\put(1426.0,368.0){\rule[-0.200pt]{2.409pt}{0.400pt}}
\put(220.0,379.0){\rule[-0.200pt]{2.409pt}{0.400pt}}
\put(1426.0,379.0){\rule[-0.200pt]{2.409pt}{0.400pt}}
\put(220.0,389.0){\rule[-0.200pt]{4.818pt}{0.400pt}}
\put(198,389){\makebox(0,0)[r]{1}}
\put(1416.0,389.0){\rule[-0.200pt]{4.818pt}{0.400pt}}
\put(220.0,453.0){\rule[-0.200pt]{2.409pt}{0.400pt}}
\put(1426.0,453.0){\rule[-0.200pt]{2.409pt}{0.400pt}}
\put(220.0,490.0){\rule[-0.200pt]{2.409pt}{0.400pt}}
\put(1426.0,490.0){\rule[-0.200pt]{2.409pt}{0.400pt}}
\put(220.0,517.0){\rule[-0.200pt]{2.409pt}{0.400pt}}
\put(1426.0,517.0){\rule[-0.200pt]{2.409pt}{0.400pt}}
\put(220.0,537.0){\rule[-0.200pt]{2.409pt}{0.400pt}}
\put(1426.0,537.0){\rule[-0.200pt]{2.409pt}{0.400pt}}
\put(220.0,554.0){\rule[-0.200pt]{2.409pt}{0.400pt}}
\put(1426.0,554.0){\rule[-0.200pt]{2.409pt}{0.400pt}}
\put(220.0,568.0){\rule[-0.200pt]{2.409pt}{0.400pt}}
\put(1426.0,568.0){\rule[-0.200pt]{2.409pt}{0.400pt}}
\put(220.0,580.0){\rule[-0.200pt]{2.409pt}{0.400pt}}
\put(1426.0,580.0){\rule[-0.200pt]{2.409pt}{0.400pt}}
\put(220.0,591.0){\rule[-0.200pt]{2.409pt}{0.400pt}}
\put(1426.0,591.0){\rule[-0.200pt]{2.409pt}{0.400pt}}
\put(220.0,601.0){\rule[-0.200pt]{4.818pt}{0.400pt}}
\put(198,601){\makebox(0,0)[r]{10}}
\put(1416.0,601.0){\rule[-0.200pt]{4.818pt}{0.400pt}}
\put(220.0,665.0){\rule[-0.200pt]{2.409pt}{0.400pt}}
\put(1426.0,665.0){\rule[-0.200pt]{2.409pt}{0.400pt}}
\put(220.0,702.0){\rule[-0.200pt]{2.409pt}{0.400pt}}
\put(1426.0,702.0){\rule[-0.200pt]{2.409pt}{0.400pt}}
\put(220.0,729.0){\rule[-0.200pt]{2.409pt}{0.400pt}}
\put(1426.0,729.0){\rule[-0.200pt]{2.409pt}{0.400pt}}
\put(220.0,749.0){\rule[-0.200pt]{2.409pt}{0.400pt}}
\put(1426.0,749.0){\rule[-0.200pt]{2.409pt}{0.400pt}}
\put(220.0,766.0){\rule[-0.200pt]{2.409pt}{0.400pt}}
\put(1426.0,766.0){\rule[-0.200pt]{2.409pt}{0.400pt}}
\put(220.0,780.0){\rule[-0.200pt]{2.409pt}{0.400pt}}
\put(1426.0,780.0){\rule[-0.200pt]{2.409pt}{0.400pt}}
\put(220.0,793.0){\rule[-0.200pt]{2.409pt}{0.400pt}}
\put(1426.0,793.0){\rule[-0.200pt]{2.409pt}{0.400pt}}
\put(220.0,803.0){\rule[-0.200pt]{2.409pt}{0.400pt}}
\put(1426.0,803.0){\rule[-0.200pt]{2.409pt}{0.400pt}}
\put(220.0,813.0){\rule[-0.200pt]{4.818pt}{0.400pt}}
\put(198,813){\makebox(0,0)[r]{100}}
\put(1416.0,813.0){\rule[-0.200pt]{4.818pt}{0.400pt}}
\put(220.0,877.0){\rule[-0.200pt]{2.409pt}{0.400pt}}
\put(1426.0,877.0){\rule[-0.200pt]{2.409pt}{0.400pt}}
\put(220.0,113.0){\rule[-0.200pt]{0.400pt}{2.409pt}}
\put(220.0,867.0){\rule[-0.200pt]{0.400pt}{2.409pt}}
\put(251.0,113.0){\rule[-0.200pt]{0.400pt}{2.409pt}}
\put(251.0,867.0){\rule[-0.200pt]{0.400pt}{2.409pt}}
\put(277.0,113.0){\rule[-0.200pt]{0.400pt}{2.409pt}}
\put(277.0,867.0){\rule[-0.200pt]{0.400pt}{2.409pt}}
\put(300.0,113.0){\rule[-0.200pt]{0.400pt}{2.409pt}}
\put(300.0,867.0){\rule[-0.200pt]{0.400pt}{2.409pt}}
\put(320.0,113.0){\rule[-0.200pt]{0.400pt}{2.409pt}}
\put(320.0,867.0){\rule[-0.200pt]{0.400pt}{2.409pt}}
\put(338.0,113.0){\rule[-0.200pt]{0.400pt}{4.818pt}}
\put(338,68){\makebox(0,0){0.01}}
\put(338.0,857.0){\rule[-0.200pt]{0.400pt}{4.818pt}}
\put(456.0,113.0){\rule[-0.200pt]{0.400pt}{2.409pt}}
\put(456.0,867.0){\rule[-0.200pt]{0.400pt}{2.409pt}}
\put(525.0,113.0){\rule[-0.200pt]{0.400pt}{2.409pt}}
\put(525.0,867.0){\rule[-0.200pt]{0.400pt}{2.409pt}}
\put(574.0,113.0){\rule[-0.200pt]{0.400pt}{2.409pt}}
\put(574.0,867.0){\rule[-0.200pt]{0.400pt}{2.409pt}}
\put(612.0,113.0){\rule[-0.200pt]{0.400pt}{2.409pt}}
\put(612.0,867.0){\rule[-0.200pt]{0.400pt}{2.409pt}}
\put(643.0,113.0){\rule[-0.200pt]{0.400pt}{2.409pt}}
\put(643.0,867.0){\rule[-0.200pt]{0.400pt}{2.409pt}}
\put(670.0,113.0){\rule[-0.200pt]{0.400pt}{2.409pt}}
\put(670.0,867.0){\rule[-0.200pt]{0.400pt}{2.409pt}}
\put(692.0,113.0){\rule[-0.200pt]{0.400pt}{2.409pt}}
\put(692.0,867.0){\rule[-0.200pt]{0.400pt}{2.409pt}}
\put(713.0,113.0){\rule[-0.200pt]{0.400pt}{2.409pt}}
\put(713.0,867.0){\rule[-0.200pt]{0.400pt}{2.409pt}}
\put(731.0,113.0){\rule[-0.200pt]{0.400pt}{4.818pt}}
\put(731,68){\makebox(0,0){0.1}}
\put(731.0,857.0){\rule[-0.200pt]{0.400pt}{4.818pt}}
\put(849.0,113.0){\rule[-0.200pt]{0.400pt}{2.409pt}}
\put(849.0,867.0){\rule[-0.200pt]{0.400pt}{2.409pt}}
\put(918.0,113.0){\rule[-0.200pt]{0.400pt}{2.409pt}}
\put(918.0,867.0){\rule[-0.200pt]{0.400pt}{2.409pt}}
\put(967.0,113.0){\rule[-0.200pt]{0.400pt}{2.409pt}}
\put(967.0,867.0){\rule[-0.200pt]{0.400pt}{2.409pt}}
\put(1005.0,113.0){\rule[-0.200pt]{0.400pt}{2.409pt}}
\put(1005.0,867.0){\rule[-0.200pt]{0.400pt}{2.409pt}}
\put(1036.0,113.0){\rule[-0.200pt]{0.400pt}{2.409pt}}
\put(1036.0,867.0){\rule[-0.200pt]{0.400pt}{2.409pt}}
\put(1062.0,113.0){\rule[-0.200pt]{0.400pt}{2.409pt}}
\put(1062.0,867.0){\rule[-0.200pt]{0.400pt}{2.409pt}}
\put(1085.0,113.0){\rule[-0.200pt]{0.400pt}{2.409pt}}
\put(1085.0,867.0){\rule[-0.200pt]{0.400pt}{2.409pt}}
\put(1105.0,113.0){\rule[-0.200pt]{0.400pt}{2.409pt}}
\put(1105.0,867.0){\rule[-0.200pt]{0.400pt}{2.409pt}}
\put(1123.0,113.0){\rule[-0.200pt]{0.400pt}{4.818pt}}
\put(1123,68){\makebox(0,0){1}}
\put(1123.0,857.0){\rule[-0.200pt]{0.400pt}{4.818pt}}
\put(1241.0,113.0){\rule[-0.200pt]{0.400pt}{2.409pt}}
\put(1241.0,867.0){\rule[-0.200pt]{0.400pt}{2.409pt}}
\put(1310.0,113.0){\rule[-0.200pt]{0.400pt}{2.409pt}}
\put(1310.0,867.0){\rule[-0.200pt]{0.400pt}{2.409pt}}
\put(1359.0,113.0){\rule[-0.200pt]{0.400pt}{2.409pt}}
\put(1359.0,867.0){\rule[-0.200pt]{0.400pt}{2.409pt}}
\put(1397.0,113.0){\rule[-0.200pt]{0.400pt}{2.409pt}}
\put(1397.0,867.0){\rule[-0.200pt]{0.400pt}{2.409pt}}
\put(1428.0,113.0){\rule[-0.200pt]{0.400pt}{2.409pt}}
\put(1428.0,867.0){\rule[-0.200pt]{0.400pt}{2.409pt}}
\put(220.0,113.0){\rule[-0.200pt]{292.934pt}{0.400pt}}
\put(1436.0,113.0){\rule[-0.200pt]{0.400pt}{184.048pt}}
\put(220.0,877.0){\rule[-0.200pt]{292.934pt}{0.400pt}}
\put(45,495){\makebox(0,0){$\Gamma(\theta)$}}
\put(828,23){\makebox(0,0){$\theta$}}
\put(220.0,113.0){\rule[-0.200pt]{0.400pt}{184.048pt}}
\put(220,731){\usebox{\plotpoint}}
\multiput(220.00,729.93)(1.267,-0.477){7}{\rule{1.060pt}{0.115pt}}
\multiput(220.00,730.17)(9.800,-5.000){2}{\rule{0.530pt}{0.400pt}}
\multiput(232.00,724.93)(1.378,-0.477){7}{\rule{1.140pt}{0.115pt}}
\multiput(232.00,725.17)(10.634,-5.000){2}{\rule{0.570pt}{0.400pt}}
\multiput(245.00,719.94)(1.651,-0.468){5}{\rule{1.300pt}{0.113pt}}
\multiput(245.00,720.17)(9.302,-4.000){2}{\rule{0.650pt}{0.400pt}}
\multiput(257.00,715.93)(1.267,-0.477){7}{\rule{1.060pt}{0.115pt}}
\multiput(257.00,716.17)(9.800,-5.000){2}{\rule{0.530pt}{0.400pt}}
\multiput(269.00,710.93)(1.267,-0.477){7}{\rule{1.060pt}{0.115pt}}
\multiput(269.00,711.17)(9.800,-5.000){2}{\rule{0.530pt}{0.400pt}}
\multiput(281.00,705.94)(1.797,-0.468){5}{\rule{1.400pt}{0.113pt}}
\multiput(281.00,706.17)(10.094,-4.000){2}{\rule{0.700pt}{0.400pt}}
\multiput(294.00,701.93)(1.267,-0.477){7}{\rule{1.060pt}{0.115pt}}
\multiput(294.00,702.17)(9.800,-5.000){2}{\rule{0.530pt}{0.400pt}}
\multiput(306.00,696.93)(1.267,-0.477){7}{\rule{1.060pt}{0.115pt}}
\multiput(306.00,697.17)(9.800,-5.000){2}{\rule{0.530pt}{0.400pt}}
\multiput(318.00,691.94)(1.797,-0.468){5}{\rule{1.400pt}{0.113pt}}
\multiput(318.00,692.17)(10.094,-4.000){2}{\rule{0.700pt}{0.400pt}}
\multiput(331.00,687.93)(1.267,-0.477){7}{\rule{1.060pt}{0.115pt}}
\multiput(331.00,688.17)(9.800,-5.000){2}{\rule{0.530pt}{0.400pt}}
\multiput(343.00,682.93)(1.267,-0.477){7}{\rule{1.060pt}{0.115pt}}
\multiput(343.00,683.17)(9.800,-5.000){2}{\rule{0.530pt}{0.400pt}}
\multiput(355.00,677.94)(1.651,-0.468){5}{\rule{1.300pt}{0.113pt}}
\multiput(355.00,678.17)(9.302,-4.000){2}{\rule{0.650pt}{0.400pt}}
\multiput(367.00,673.93)(1.378,-0.477){7}{\rule{1.140pt}{0.115pt}}
\multiput(367.00,674.17)(10.634,-5.000){2}{\rule{0.570pt}{0.400pt}}
\multiput(380.00,668.94)(1.651,-0.468){5}{\rule{1.300pt}{0.113pt}}
\multiput(380.00,669.17)(9.302,-4.000){2}{\rule{0.650pt}{0.400pt}}
\multiput(392.00,664.93)(1.267,-0.477){7}{\rule{1.060pt}{0.115pt}}
\multiput(392.00,665.17)(9.800,-5.000){2}{\rule{0.530pt}{0.400pt}}
\multiput(404.00,659.93)(1.378,-0.477){7}{\rule{1.140pt}{0.115pt}}
\multiput(404.00,660.17)(10.634,-5.000){2}{\rule{0.570pt}{0.400pt}}
\multiput(417.00,654.94)(1.651,-0.468){5}{\rule{1.300pt}{0.113pt}}
\multiput(417.00,655.17)(9.302,-4.000){2}{\rule{0.650pt}{0.400pt}}
\multiput(429.00,650.93)(1.267,-0.477){7}{\rule{1.060pt}{0.115pt}}
\multiput(429.00,651.17)(9.800,-5.000){2}{\rule{0.530pt}{0.400pt}}
\multiput(441.00,645.93)(1.267,-0.477){7}{\rule{1.060pt}{0.115pt}}
\multiput(441.00,646.17)(9.800,-5.000){2}{\rule{0.530pt}{0.400pt}}
\multiput(453.00,640.94)(1.797,-0.468){5}{\rule{1.400pt}{0.113pt}}
\multiput(453.00,641.17)(10.094,-4.000){2}{\rule{0.700pt}{0.400pt}}
\multiput(466.00,636.93)(1.267,-0.477){7}{\rule{1.060pt}{0.115pt}}
\multiput(466.00,637.17)(9.800,-5.000){2}{\rule{0.530pt}{0.400pt}}
\multiput(478.00,631.93)(1.267,-0.477){7}{\rule{1.060pt}{0.115pt}}
\multiput(478.00,632.17)(9.800,-5.000){2}{\rule{0.530pt}{0.400pt}}
\multiput(490.00,626.94)(1.797,-0.468){5}{\rule{1.400pt}{0.113pt}}
\multiput(490.00,627.17)(10.094,-4.000){2}{\rule{0.700pt}{0.400pt}}
\multiput(503.00,622.93)(1.267,-0.477){7}{\rule{1.060pt}{0.115pt}}
\multiput(503.00,623.17)(9.800,-5.000){2}{\rule{0.530pt}{0.400pt}}
\multiput(515.00,617.93)(1.267,-0.477){7}{\rule{1.060pt}{0.115pt}}
\multiput(515.00,618.17)(9.800,-5.000){2}{\rule{0.530pt}{0.400pt}}
\multiput(527.00,612.94)(1.651,-0.468){5}{\rule{1.300pt}{0.113pt}}
\multiput(527.00,613.17)(9.302,-4.000){2}{\rule{0.650pt}{0.400pt}}
\multiput(539.00,608.93)(1.378,-0.477){7}{\rule{1.140pt}{0.115pt}}
\multiput(539.00,609.17)(10.634,-5.000){2}{\rule{0.570pt}{0.400pt}}
\multiput(552.00,603.93)(1.267,-0.477){7}{\rule{1.060pt}{0.115pt}}
\multiput(552.00,604.17)(9.800,-5.000){2}{\rule{0.530pt}{0.400pt}}
\multiput(564.00,598.94)(1.651,-0.468){5}{\rule{1.300pt}{0.113pt}}
\multiput(564.00,599.17)(9.302,-4.000){2}{\rule{0.650pt}{0.400pt}}
\multiput(576.00,594.93)(1.267,-0.477){7}{\rule{1.060pt}{0.115pt}}
\multiput(576.00,595.17)(9.800,-5.000){2}{\rule{0.530pt}{0.400pt}}
\multiput(588.00,589.94)(1.797,-0.468){5}{\rule{1.400pt}{0.113pt}}
\multiput(588.00,590.17)(10.094,-4.000){2}{\rule{0.700pt}{0.400pt}}
\multiput(601.00,585.93)(1.267,-0.477){7}{\rule{1.060pt}{0.115pt}}
\multiput(601.00,586.17)(9.800,-5.000){2}{\rule{0.530pt}{0.400pt}}
\multiput(613.00,580.93)(1.267,-0.477){7}{\rule{1.060pt}{0.115pt}}
\multiput(613.00,581.17)(9.800,-5.000){2}{\rule{0.530pt}{0.400pt}}
\multiput(625.00,575.94)(1.797,-0.468){5}{\rule{1.400pt}{0.113pt}}
\multiput(625.00,576.17)(10.094,-4.000){2}{\rule{0.700pt}{0.400pt}}
\multiput(638.00,571.93)(1.267,-0.477){7}{\rule{1.060pt}{0.115pt}}
\multiput(638.00,572.17)(9.800,-5.000){2}{\rule{0.530pt}{0.400pt}}
\multiput(650.00,566.93)(1.267,-0.477){7}{\rule{1.060pt}{0.115pt}}
\multiput(650.00,567.17)(9.800,-5.000){2}{\rule{0.530pt}{0.400pt}}
\multiput(662.00,561.94)(1.651,-0.468){5}{\rule{1.300pt}{0.113pt}}
\multiput(662.00,562.17)(9.302,-4.000){2}{\rule{0.650pt}{0.400pt}}
\multiput(674.00,557.93)(1.378,-0.477){7}{\rule{1.140pt}{0.115pt}}
\multiput(674.00,558.17)(10.634,-5.000){2}{\rule{0.570pt}{0.400pt}}
\multiput(687.00,552.93)(1.267,-0.477){7}{\rule{1.060pt}{0.115pt}}
\multiput(687.00,553.17)(9.800,-5.000){2}{\rule{0.530pt}{0.400pt}}
\multiput(699.00,547.94)(1.651,-0.468){5}{\rule{1.300pt}{0.113pt}}
\multiput(699.00,548.17)(9.302,-4.000){2}{\rule{0.650pt}{0.400pt}}
\multiput(711.00,543.93)(1.378,-0.477){7}{\rule{1.140pt}{0.115pt}}
\multiput(711.00,544.17)(10.634,-5.000){2}{\rule{0.570pt}{0.400pt}}
\multiput(724.00,538.93)(1.267,-0.477){7}{\rule{1.060pt}{0.115pt}}
\multiput(724.00,539.17)(9.800,-5.000){2}{\rule{0.530pt}{0.400pt}}
\multiput(736.00,533.94)(1.651,-0.468){5}{\rule{1.300pt}{0.113pt}}
\multiput(736.00,534.17)(9.302,-4.000){2}{\rule{0.650pt}{0.400pt}}
\multiput(748.00,529.93)(1.267,-0.477){7}{\rule{1.060pt}{0.115pt}}
\multiput(748.00,530.17)(9.800,-5.000){2}{\rule{0.530pt}{0.400pt}}
\multiput(760.00,524.94)(1.797,-0.468){5}{\rule{1.400pt}{0.113pt}}
\multiput(760.00,525.17)(10.094,-4.000){2}{\rule{0.700pt}{0.400pt}}
\multiput(773.00,520.93)(1.267,-0.477){7}{\rule{1.060pt}{0.115pt}}
\multiput(773.00,521.17)(9.800,-5.000){2}{\rule{0.530pt}{0.400pt}}
\multiput(785.00,515.93)(1.267,-0.477){7}{\rule{1.060pt}{0.115pt}}
\multiput(785.00,516.17)(9.800,-5.000){2}{\rule{0.530pt}{0.400pt}}
\multiput(797.00,510.94)(1.797,-0.468){5}{\rule{1.400pt}{0.113pt}}
\multiput(797.00,511.17)(10.094,-4.000){2}{\rule{0.700pt}{0.400pt}}
\multiput(810.00,506.93)(1.267,-0.477){7}{\rule{1.060pt}{0.115pt}}
\multiput(810.00,507.17)(9.800,-5.000){2}{\rule{0.530pt}{0.400pt}}
\multiput(822.00,501.93)(1.267,-0.477){7}{\rule{1.060pt}{0.115pt}}
\multiput(822.00,502.17)(9.800,-5.000){2}{\rule{0.530pt}{0.400pt}}
\multiput(834.00,496.94)(1.651,-0.468){5}{\rule{1.300pt}{0.113pt}}
\multiput(834.00,497.17)(9.302,-4.000){2}{\rule{0.650pt}{0.400pt}}
\multiput(846.00,492.93)(1.378,-0.477){7}{\rule{1.140pt}{0.115pt}}
\multiput(846.00,493.17)(10.634,-5.000){2}{\rule{0.570pt}{0.400pt}}
\multiput(859.00,487.94)(1.651,-0.468){5}{\rule{1.300pt}{0.113pt}}
\multiput(859.00,488.17)(9.302,-4.000){2}{\rule{0.650pt}{0.400pt}}
\multiput(871.00,483.93)(1.267,-0.477){7}{\rule{1.060pt}{0.115pt}}
\multiput(871.00,484.17)(9.800,-5.000){2}{\rule{0.530pt}{0.400pt}}
\multiput(883.00,478.93)(1.378,-0.477){7}{\rule{1.140pt}{0.115pt}}
\multiput(883.00,479.17)(10.634,-5.000){2}{\rule{0.570pt}{0.400pt}}
\multiput(896.00,473.94)(1.651,-0.468){5}{\rule{1.300pt}{0.113pt}}
\multiput(896.00,474.17)(9.302,-4.000){2}{\rule{0.650pt}{0.400pt}}
\multiput(908.00,469.93)(1.267,-0.477){7}{\rule{1.060pt}{0.115pt}}
\multiput(908.00,470.17)(9.800,-5.000){2}{\rule{0.530pt}{0.400pt}}
\multiput(920.00,464.94)(1.651,-0.468){5}{\rule{1.300pt}{0.113pt}}
\multiput(920.00,465.17)(9.302,-4.000){2}{\rule{0.650pt}{0.400pt}}
\multiput(932.00,460.93)(1.378,-0.477){7}{\rule{1.140pt}{0.115pt}}
\multiput(932.00,461.17)(10.634,-5.000){2}{\rule{0.570pt}{0.400pt}}
\multiput(945.00,455.94)(1.651,-0.468){5}{\rule{1.300pt}{0.113pt}}
\multiput(945.00,456.17)(9.302,-4.000){2}{\rule{0.650pt}{0.400pt}}
\multiput(957.00,451.93)(1.267,-0.477){7}{\rule{1.060pt}{0.115pt}}
\multiput(957.00,452.17)(9.800,-5.000){2}{\rule{0.530pt}{0.400pt}}
\multiput(969.00,446.94)(1.797,-0.468){5}{\rule{1.400pt}{0.113pt}}
\multiput(969.00,447.17)(10.094,-4.000){2}{\rule{0.700pt}{0.400pt}}
\multiput(982.00,442.93)(1.267,-0.477){7}{\rule{1.060pt}{0.115pt}}
\multiput(982.00,443.17)(9.800,-5.000){2}{\rule{0.530pt}{0.400pt}}
\multiput(994.00,437.94)(1.651,-0.468){5}{\rule{1.300pt}{0.113pt}}
\multiput(994.00,438.17)(9.302,-4.000){2}{\rule{0.650pt}{0.400pt}}
\multiput(1006.00,433.93)(1.267,-0.477){7}{\rule{1.060pt}{0.115pt}}
\multiput(1006.00,434.17)(9.800,-5.000){2}{\rule{0.530pt}{0.400pt}}
\multiput(1018.00,428.94)(1.797,-0.468){5}{\rule{1.400pt}{0.113pt}}
\multiput(1018.00,429.17)(10.094,-4.000){2}{\rule{0.700pt}{0.400pt}}
\multiput(1031.00,424.94)(1.651,-0.468){5}{\rule{1.300pt}{0.113pt}}
\multiput(1031.00,425.17)(9.302,-4.000){2}{\rule{0.650pt}{0.400pt}}
\multiput(1043.00,420.93)(1.267,-0.477){7}{\rule{1.060pt}{0.115pt}}
\multiput(1043.00,421.17)(9.800,-5.000){2}{\rule{0.530pt}{0.400pt}}
\multiput(1055.00,415.94)(1.797,-0.468){5}{\rule{1.400pt}{0.113pt}}
\multiput(1055.00,416.17)(10.094,-4.000){2}{\rule{0.700pt}{0.400pt}}
\multiput(1068.00,411.94)(1.651,-0.468){5}{\rule{1.300pt}{0.113pt}}
\multiput(1068.00,412.17)(9.302,-4.000){2}{\rule{0.650pt}{0.400pt}}
\multiput(1080.00,407.94)(1.651,-0.468){5}{\rule{1.300pt}{0.113pt}}
\multiput(1080.00,408.17)(9.302,-4.000){2}{\rule{0.650pt}{0.400pt}}
\multiput(1092.00,403.94)(1.651,-0.468){5}{\rule{1.300pt}{0.113pt}}
\multiput(1092.00,404.17)(9.302,-4.000){2}{\rule{0.650pt}{0.400pt}}
\multiput(1104.00,399.94)(1.797,-0.468){5}{\rule{1.400pt}{0.113pt}}
\multiput(1104.00,400.17)(10.094,-4.000){2}{\rule{0.700pt}{0.400pt}}
\multiput(1117.00,395.95)(2.472,-0.447){3}{\rule{1.700pt}{0.108pt}}
\multiput(1117.00,396.17)(8.472,-3.000){2}{\rule{0.850pt}{0.400pt}}
\multiput(1129.00,392.94)(1.651,-0.468){5}{\rule{1.300pt}{0.113pt}}
\multiput(1129.00,393.17)(9.302,-4.000){2}{\rule{0.650pt}{0.400pt}}
\multiput(1141.00,388.95)(2.472,-0.447){3}{\rule{1.700pt}{0.108pt}}
\multiput(1141.00,389.17)(8.472,-3.000){2}{\rule{0.850pt}{0.400pt}}
\multiput(1153.00,385.95)(2.695,-0.447){3}{\rule{1.833pt}{0.108pt}}
\multiput(1153.00,386.17)(9.195,-3.000){2}{\rule{0.917pt}{0.400pt}}
\multiput(1166.00,382.95)(2.472,-0.447){3}{\rule{1.700pt}{0.108pt}}
\multiput(1166.00,383.17)(8.472,-3.000){2}{\rule{0.850pt}{0.400pt}}
\multiput(1178.00,379.95)(2.472,-0.447){3}{\rule{1.700pt}{0.108pt}}
\multiput(1178.00,380.17)(8.472,-3.000){2}{\rule{0.850pt}{0.400pt}}
\put(1190,376.17){\rule{2.700pt}{0.400pt}}
\multiput(1190.00,377.17)(7.396,-2.000){2}{\rule{1.350pt}{0.400pt}}
\put(1203,374.17){\rule{2.500pt}{0.400pt}}
\multiput(1203.00,375.17)(6.811,-2.000){2}{\rule{1.250pt}{0.400pt}}
\put(1215,372.67){\rule{2.891pt}{0.400pt}}
\multiput(1215.00,373.17)(6.000,-1.000){2}{\rule{1.445pt}{0.400pt}}
\put(1239,372.67){\rule{3.132pt}{0.400pt}}
\multiput(1239.00,372.17)(6.500,1.000){2}{\rule{1.566pt}{0.400pt}}
\multiput(1252.00,374.61)(2.472,0.447){3}{\rule{1.700pt}{0.108pt}}
\multiput(1252.00,373.17)(8.472,3.000){2}{\rule{0.850pt}{0.400pt}}
\multiput(1264.00,377.60)(1.651,0.468){5}{\rule{1.300pt}{0.113pt}}
\multiput(1264.00,376.17)(9.302,4.000){2}{\rule{0.650pt}{0.400pt}}
\multiput(1276.00,381.59)(0.824,0.488){13}{\rule{0.750pt}{0.117pt}}
\multiput(1276.00,380.17)(11.443,8.000){2}{\rule{0.375pt}{0.400pt}}
\multiput(1289.58,389.00)(0.492,0.669){21}{\rule{0.119pt}{0.633pt}}
\multiput(1288.17,389.00)(12.000,14.685){2}{\rule{0.400pt}{0.317pt}}
\multiput(1301.58,405.00)(0.492,1.832){21}{\rule{0.119pt}{1.533pt}}
\multiput(1300.17,405.00)(12.000,39.817){2}{\rule{0.400pt}{0.767pt}}
\put(1227.0,373.0){\rule[-0.200pt]{2.891pt}{0.400pt}}
\end{picture}
\end{figure}

\begin{thebibliography}{Mandelbrot}       
\bibitem{Borgani} Borgani S 1995 {\em Phys. Reports} {\bf 251} 1 
\bibitem{Davis}  Davis M 1996 {\em 
Round table discussion at the conference
``Critical Dialogues in Cosmology''} (Princeton);
LANL e-print astro-ph/9610149
\bibitem{Pietronero} Pietronero L,  Montuori M and  Sylos Labini F
 1996 {Round table discussion at the conference
``Critical Dialogues in Cosmology''} (Princeton); LANL e-print
 astro-ph/9611197
\bibitem{CP}  Coleman  P H and  Pietronero L 1992
 {\em Phys. Reports}  {\bf 213} 311, and references therein
\bibitem{Shi} Shi Y 1997  (submitted)
\end{thebibliography}
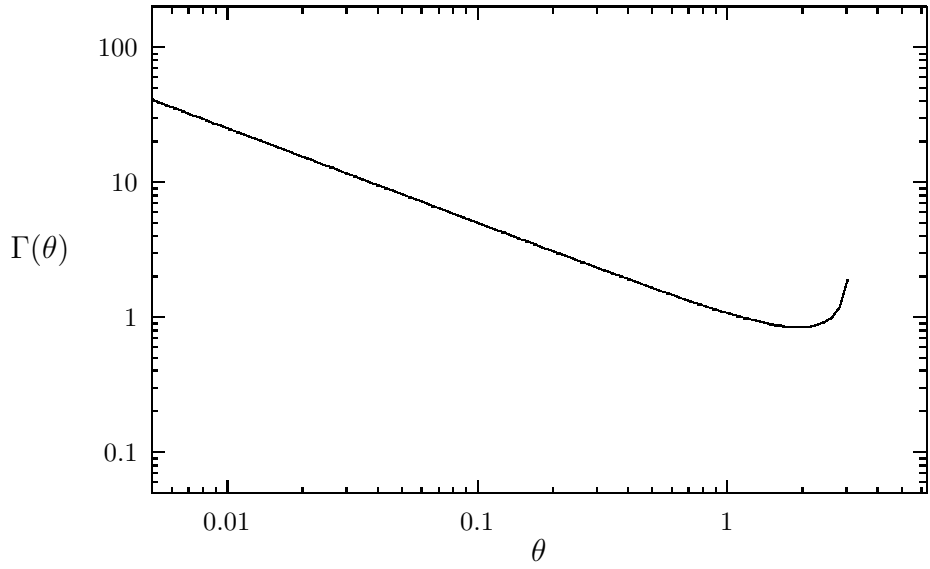
\end{document}